# Very high critical field and superior $J_c$-field performance in NdFeAsO$_{0.82}$F$_{0.18}$ with $T_c$ of 51 K

X.L. Wang[1]*, R. Ghorbani[1,2], G. Peleckis[1], S.X. Dou[1]*

[1]*Institute for Superconducting and Electronic Materials, University of Wollongong, Wollongong, NSW 2522, Australia*

[2]*Department of Physics, Tarbiat Moallem University of Sabzevar, P.O. Box 397, Sabzevar, Iran*

The discovery of the new family of oxypnictide superconductors [1,2], including LaFeAsO$_{0.89}$F$_{0.11}$, with critical temperature, $T_c$, over 26 K, has brought new impetus to the fields of high temperature superconductivity and strongly correlated electron systems. The new superconductors have the general formula REFeAsO, where RE is a rare earth element [1-8], and show two dimensional crystal structures similar to those of the high $T_c$ cuprate superconductors. They consist of alternating REO and FeAs layers, providing charge carriers and conducting planes, respectively. The $T_c$ is strongly dependent on the sizes of the rare earth element ions [1-8], as well as on F doping on oxygen sites [1,2] and oxygen deficiency in F free material [4]. The upper critical field, $H_{c2}$, has been estimated to be higher than 55 or 63-65 T in LaFeAsO$_{0.9}$F$_{0.1}$ [6,7], with $T_c$ of 26 K, 70 T in PrFeAsO$_{0.85}$F$_{0.15}$, and over 100 T in SmFeAsO$_{0.85}$F$_{0.15}$ [7], with $T_c$ of 46 K. The two gap superconductivity proposed for LaFeAsO$_{0.9}$F$_{0.1}$ suggests that the $H_{c2}$ could be further increased [6] to a significant extent. This is one of the unique features of the FeAs-based new superconductors. In this work, we show that the $H_{c2}$(48 K) = 13 T, and the $H_{c2}$(0) values can exceed 80-230 T in a high-pressure fabricated NdO$_{0.82}$F$_{0.18}$FeAs bulk sample with $T_c$ of 51 K. We also demonstrate that the supercurrent density in fields from 1 up to 9 T only drops by a factor of 2-6 for T < 30 K, much more slowly than for MgB$_2$ and high $T_c$ cuprate superconductors. The very high $H_{c2}$ far surpasses those of MgB$_2$ and classical low temperature superconductors, and the superior $J_c$-field performance is promising for the use of the new NdFeAsO$_{0.82}$F$_{0.18}$ superconductors in high-field applications.

*Correspondence should be addressed to xiaolin@uow.edu.au or shi@uow.edu.au

A polycrystalline sample with the nominal composition NdFeAsO$_{0.82}$F$_{0.18}$ was prepared by a high-pressure (HP) technique. Powders of NdAs, As, Fe, Fe$_2$O$_3$, and FeF$_2$ were well mixed, pelletized, and then sealed in a boron nitride crucible and sintered at 1250 °C for 2 hours under the high pressure of 6 GPa [3]. The phases in the sample were investigated by powder X-ray diffraction (XRD), and the structure was refined using Rietveld refinement. Standard four-probe resistivity measurements were carried out on a bar sample by using a physical properties measurement system (PPMS, Quantum Design) in the field range from 0 up to 13 T. Magnetic loops were also collected at various temperatures below $T_c$. The critical current density was calculated by using the Bean approximation.

The XRD and refinement results shown in Fig. 1 indicate that the NdFeAsO$_{0.82}$F$_{0.18}$ sample is nearly single phase, with a tetragonal structure of the *P4/nmm* symmetry and the lattice parameters $a$ = 3.953 Å and $c$ = 8.527 Å. The Nd and As are located at the Wyckoff position 2c with $z$ = 0.143 and 0.659, respectively. The nearest neighbor distances are $d$ (Nd-O) = 2.321 Å, $d$ (Fe-As) = 2.397 Å, $d$ (La-As) = 3.268 Å, and $d$ (Fe-Fe) = 2.795 Å. These lattice parameters and bond lengths are smaller than those of LaFeAsO$_{0.89}$F$_{0.11}$, due to the fact that the size of the Nd$^{3+}$ ion is smaller than that of the La$^{3+}$. A tiny amount of Nd$_2$O$_3$ (less than 1.7 wt%) is present as the second phase.

The temperature dependence of the resistivity of the NdFeAsO$_{0.82}$F$_{0.18}$ is shown in Fig. 2. The resistivity is about 9 mΩcm at 300 K and 3 mΩcm at 52 K, while the residual resistivity ratio RRR = $\rho$(300K)/$\rho$(52 K) = 3, which means that the scattering becomes large at the onset temperature. The resistance drops to zero at $T$ = 46 K in zero magnetic field. It can been seen that the onset $T_c$ drops very slowly with increasing magnetic field, however, the $T_c$(0) decreases quickly to low temperatures. The upper critical field, $H_{c2}$, is defined as the field at which the resistance starts to drop. We use a criterion of 99% of normal resistivity at the onset temperature. The $H_{c2}$ defined this way refers to the case of field parallel to the *ab*-plane, $H_{c2}^{ab}$.

The magnetoresistance $R(B)$ was also measured at several temperatures, as shown in Fig. 3. The broad $R(B)$ transition is similar to what has been seen in a LaFeAsO$_{0.82}$F$_{0.18}$ sample [6]. Using the same analysis used in Ref. [6], the $B_{max}$, the midpoint transition field ($B_{mid}$), and $B_{min}$ are defined as 90%, 50%, and 10% of the normal state resistance $R_n(T_c)$, respectively. The magnetoresistance plots enabled us to extract some high field data by extrapolation of $R(B)$ at $B < 13$ T to $R(B) = 0.9\ R_n(T_c)$, $0.5\ R_n(T_c)$, and $0.1\ R_n(T_c)$, as shown by the dashed lines in Fig. 3. We can also define $B_{max}$ (90% $R_n$), the midpoint of the transition, $B_{mid}$ (50% $R_n$), and $B_{min}$ (10% $R_n$) from the R-T curves. The magnetic loop measurements indicate that the magnetization loops are almost reversible at T > 20 K and B < 8.7 T. This clearly implies that the $B_{max}(T)$ and $B_{min}(T)$ from both the R-T and $B(H)$ curves can be regarded as the temperature dependences of $H_{c2}^{ab}$ and $H_{c2}^{c}$. All these defined fields are plotted in Fig. 4 as a comparison.

The slope of $H_{c2}^{ab}$ near $T_c$ (99% $R_n$ from the R-T curves), i.e. $dH_{c2}^{ab}/dT$, is -5.8 T/K. This value obtained from our NdFeAsO$_{0.82}$F$_{0.18}$ sample is larger than that for LaFeAsO$_{0.89}$F$_{0.11}$ ($dH_{c2}/dT \approx 2$ T/K) and SmFeAsO$_{0.85}$F$_{0.15}$ ($dH_{c2}/dT \approx 5$ T/K) [5]. The estimated slopes $dH_{c2}^{ab}/dT$ for $H_{max}$ (90% $R_n$) and $dH_{c2}^{c}/dT$ for $H_{min}$ (10% $R_n$) are 5.6 and 2.5 T/K, respectively.

Below $B_{min}$, a low measurement current density of 0.07 A/cm$^2$ can flow through the sample via a percolative path. The $H_{c2}^{ab}(0)$ can be estimated using the Werthamer-Helfand-Hohenberg (WHH) formula: $H_{c2}^{ab}(0) = -0.69 T_c (dH_{c2}^{ab}/dT)_{T=Tc} = 204$ T for the field at 99% $R_n$ and 195 T for 90% $R_n$. $H_{c2}^{c}(0) = -0.69 T_c (dH_{c2}^{c}/dT)_{T=Tc} = 80$ T. Using the Ginzburg-Landau (GL) equation:

$$H_{c2}(T) = H_{c2}(0)\frac{1-t^2}{1+t^2} \qquad (1)$$

where $t = T/T_c$ is the reduced temperature and $H_{c2}(0)$ is the upper critical field at zero temperature. Fig. 4(c) shows the good fit of the G-L theory to the experimental data for the high temperature range. $H_{c2}^{ab}(0)$, $H_{c2}^{ab}$ (midpoint), $H_{c2}^{ab}$ (90%), and $H_{c2}^{ab}$ (99%) are estimated to be 80, 150, 230, and 310 T, respectively. It should be pointed out that although the WHH and GL equations may not be valid for the low temperature range, the $H_{c2}$ values estimated using both equations are usually far below the real experimental data. According to Ref [6], the real $H_{c2}^{ab}$ values are at least 3-6 times higher than what were estimated from the WHH law, as seen in LaFeAsO$_{0.9}$F$_{0.1}$ [6]. These indicate that the real $H_{c2}$ in the NdFeAsO$_{0.82}$F$_{0.18}$ sample should be higher than what we estimated from the WHH and GL equations. Furthermore, a strong paramagnetic state was observed and became dominant in high magnetic fields below $T_c$, as seen in a NdFeAsO$_{0.89}$F$_{0.11}$ sample [10]. This makes it difficult to estimate the $H_{c2}$ from the magnetization measurements. Therefore, we believe that the determination of the $H_{c2}$ using the transport method, as shown in this work, is more convenient and reliable compared to magnetization measurements.

According to our data, the estimated anisotropy for the NdFeAsO$_{0.82}$F$_{0.18}$ is $\Gamma = (H_{c2}^{ab}/H_{c2}^{c})^2 = 15$ and 8.3 for $H_{c2}^{ab}$ (99%) and $H_{c2}^{ab}$ (90%), respectively, which is very much in agreement with what has been predicted from the resistivity ratio, $\Gamma = \rho_c/\rho_{ab}$ of 10-15 for LaFeAsO [10].

The Ginzburg–Landau (GL) formula for the coherence length ($\xi$) is $\xi = (\Phi_0/2\pi\mu_0 H_{c2})^{1/2}$, where $\Phi_0 = 2.07 \times 10^{-7}$ Oe.cm$^2$. This yields a zero temperature coherence length, $\xi$, of 10.3 Å for $H_{c2}$(99% $R_n$) (and 14.5 Å for $H_{c2}$ at 50% $R_n$, and 12 Å for $H_{c2}$(90% $R_n$)). These values are much smaller than that reported for LaFeAsO$_{0.89}$F$_{0.11}$ (35 Å) [10] and comparable to those for high-temperature superconducting cuprates.

As a comparison, the $H_{c2}^{ab}$ and $H_{c2}^{c}$ for MgB$_2$ ($T_c$ = 39 K), LaFeAsO$_{0.89}$F$_{0.11}$ ($T_c$ = 27 K), and SmFeAsO$_{0.89}$F$_{0.11}$ ($T_c$ = 45 K) are also plotted as a function of real and reduced temperature ($T/T_c$), as shown in Fig. 4(a) and (b). It can be seen that the NdFeAsO$_{0.82}$F$_{0.18}$ sample shows the largest slopes and highest values of both $H_{c2}^{ab}$ and $H_{c2}^{c}$.

The typical ferromagnetic loops which have been seen for $T < T_c$ and $T > T_c$ in Sm-based FeAs samples [7] are absent in the NdFeAsO$_{0.82}$F$_{0.18}$ sample. This result rules out the presence of ferromagnetic impurities, such as Fe or Fe$_2$O$_3$. The magnetization loops show a strong paramagnetic background, which is believed to be contributed by the paramagnetic state from Nd$^{3+}$ [10]. According to the magneto-optical imaging studies, the intragrain critical current density is about $6 \times 10^4$ A/cm$^2$ at 20 K at low field for the NdFeAsO$_{0.82}$F$_{0.18}$ prepared using the HP method. We calculated the $J_c$ for our samples on the basis of $J_c = 20\Delta M/a(1-a/3b)$, where $\Delta M$ is the height of the magnetization loop and $a$ and $b$ are the dimensions of the sample perpendicular to the magnetic field, $a < b$, using the size of the whole sample and the average size of grains, assuming that the current flows only within grains. ($J_c = 3\Delta M/<R>$, where $<R>$ is the average grain size.) The $J_c$ field dependence is shown in Fig. 5. It can be seen that the $J_c$ based on the sample size is above $10^4$ A/cm$^2$, which is lower than what should exist in individual grains. However, the $J_c$ based on the individual grains (assuming that the average grain size is about 50-100 µm [12]) is about $2 \times 10^5 - 1 \times 10^6$ A/cm$^2$ at 5 K, which is agreement with the result obtained from the magneto-optical imaging method ($J_c \approx 2.8 \times 10^5$ A/cm). It should be noted that the $J_c$ has a very weak dependence on field and remains constant for $B > 2$ T and T = 5 K. The $J_c$ only drops less than one and two orders of magnitude up to high fields of 8 T for 5 K and 20-30 K, respectively. Furthermore, for $B > 1$ T, the $J_c$ only decreases by a factor of 2-6 for T < 30 K. These results indicate that the NdO$_{0.82}$F$_{0.18}$FeAs exhibits a superior $J_c$ - field performance compared to that of MgB$_2$, where $J_c$ drops very quickly at 20 and 30 K, even at low fields.

In summary, we have shown that NdFeAsO$_{0.82}$F$_{0.18}$ superconductor fabricated using a high pressure technique exhibits a weak $J_c$-field dependence. The upper critical field $H_{c2}$ (48 K) = 13 T, and the $H_{c2}$(0) value can exceed 80 - 230 T, which surpasses the $H_{c2}$ of all the low temperature superconductors and of MgB$_2$, and is as high as for high $T_c$ cuprate superconductors. Such a high $H_{c2}$ and the weak $J_c$-field dependence make NdFeAsO superconductor a powerful competitor that will be potentially useful in very high field applications. The $H_{c2}$ still has the potential to be enhanced by proper chemical doping, physical approaches, and other rare earth doping, due to the two-gap superconductivity in the new FeAs based superconductor. The challenge now relates to how the materials can be made with perfect texture in order to allow this new superconductor to carry a high enough critical current density.

**Acknowledgements**

The authors thank the Australian Research Council for funding support through Discovery projects. We would like to thank Profs. Z. A. Ren and Z. X. Zhao from the Institute of Physics, Chinese Academy of Sciences for providing the sample used in this work.

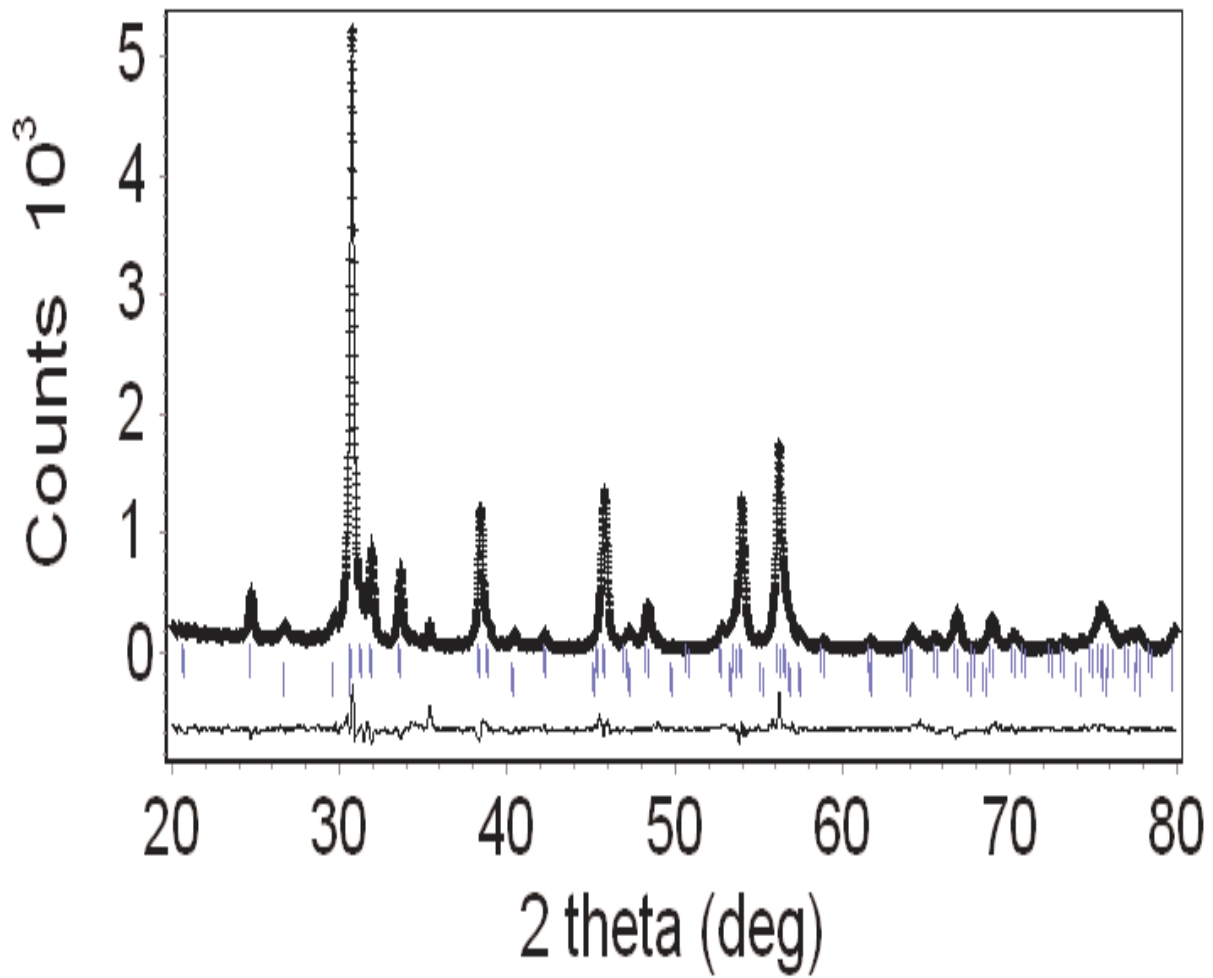

Fig. 1. The observed (crosses), calculated (solid line), and difference diffraction (bottom solid line) profiles at 300 K for NdFeAsO$_{0.82}$F$_{0.18}$. The top peak markers relate to NdFeAsO, while the lower peak markers relate to the impurity Nd$_2$O$_3$. The peak at 35 degrees is characteristic of boron nitride.

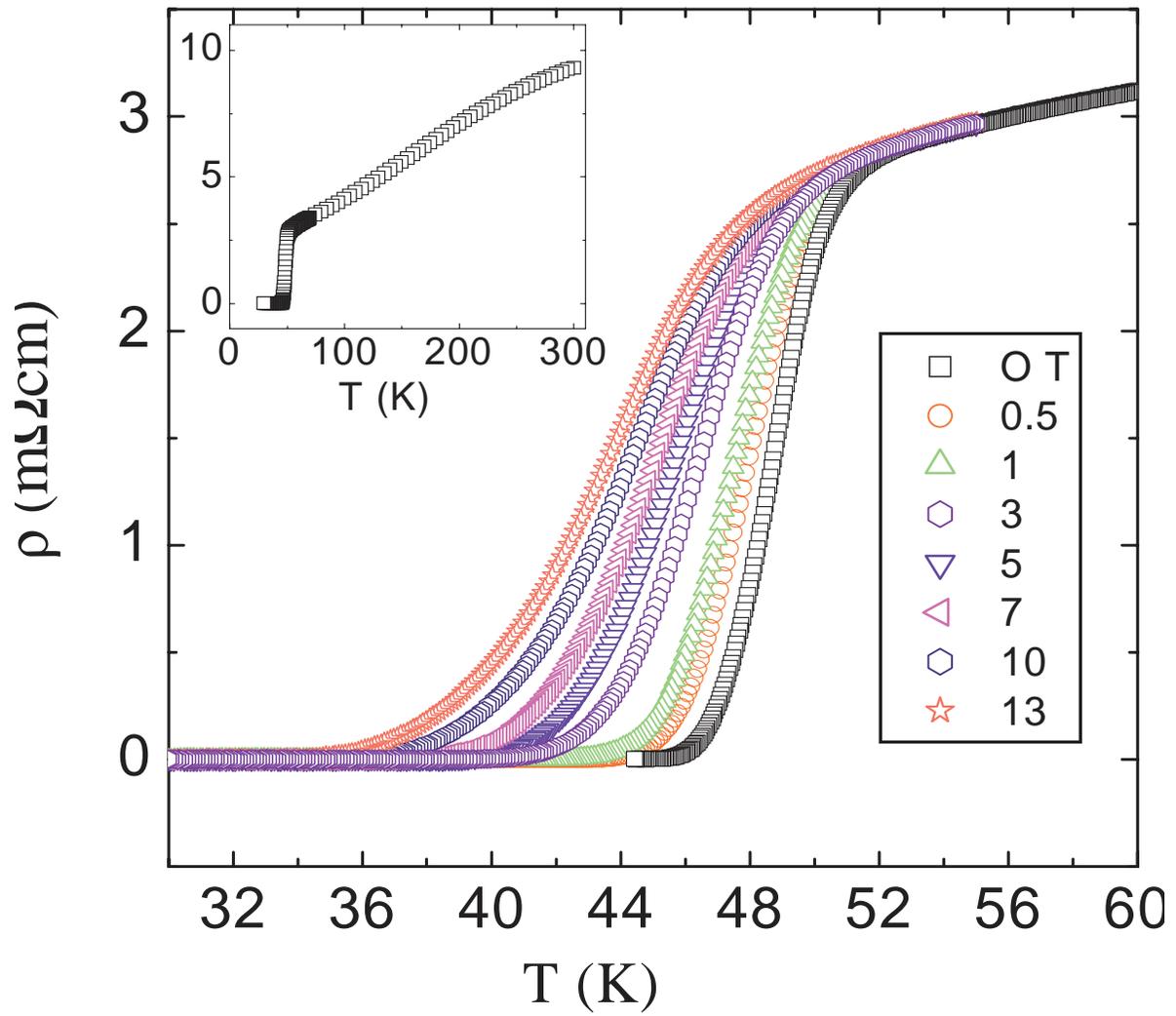

Fig. 2. The temperature dependence of the resistivity of NdFeAsO$_{0.82}$F$_{0.18}$, measured in fields up to 13 T. The inset shows the ρ-T curve from 5 up to 300 K.

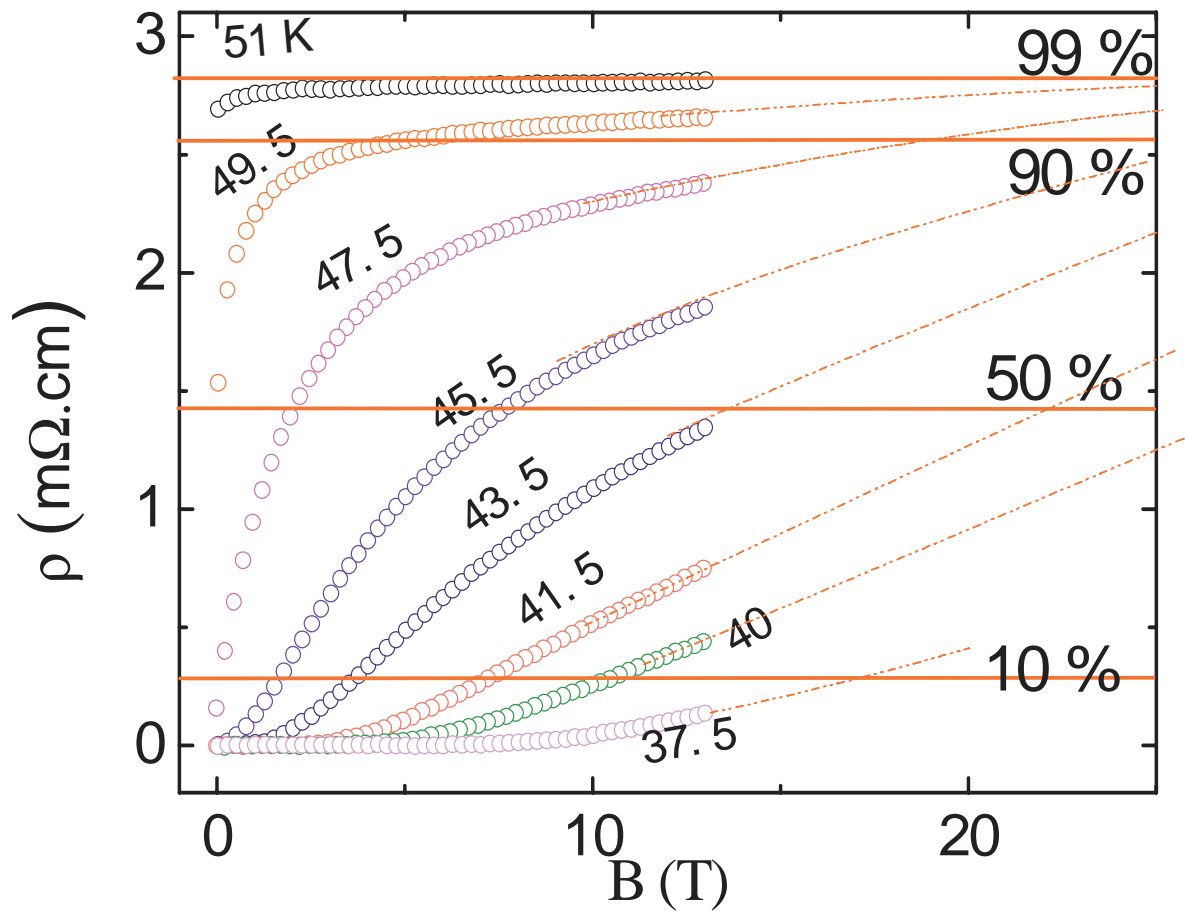

Fig. 3. The resistivity $\rho(B)$ for different temperatures, taken in swept fields in a 14 T magnet with a measurement current of 1 mA.

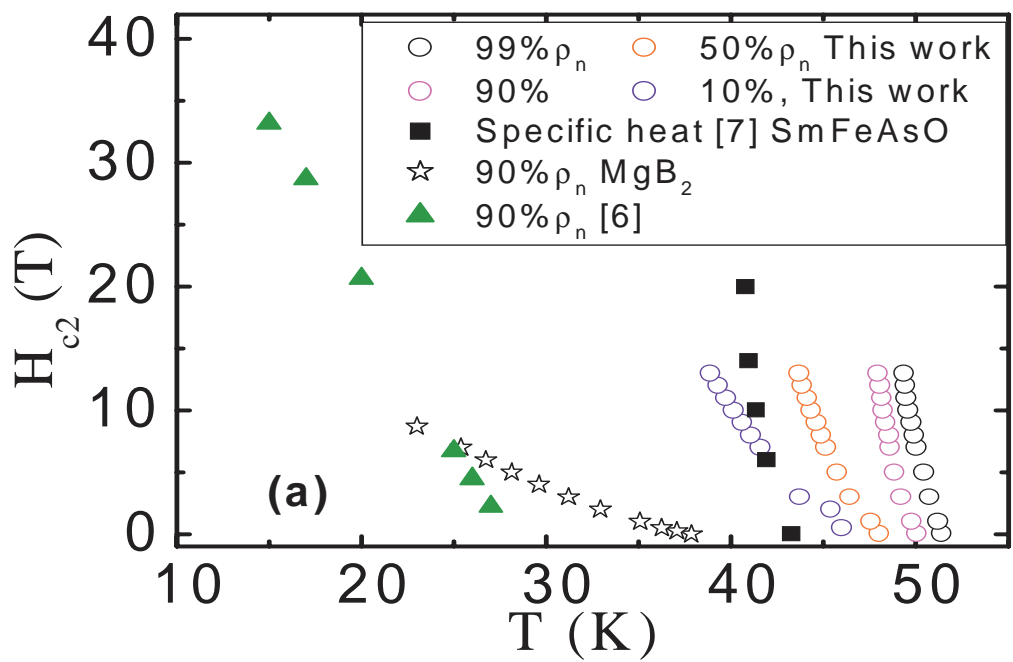

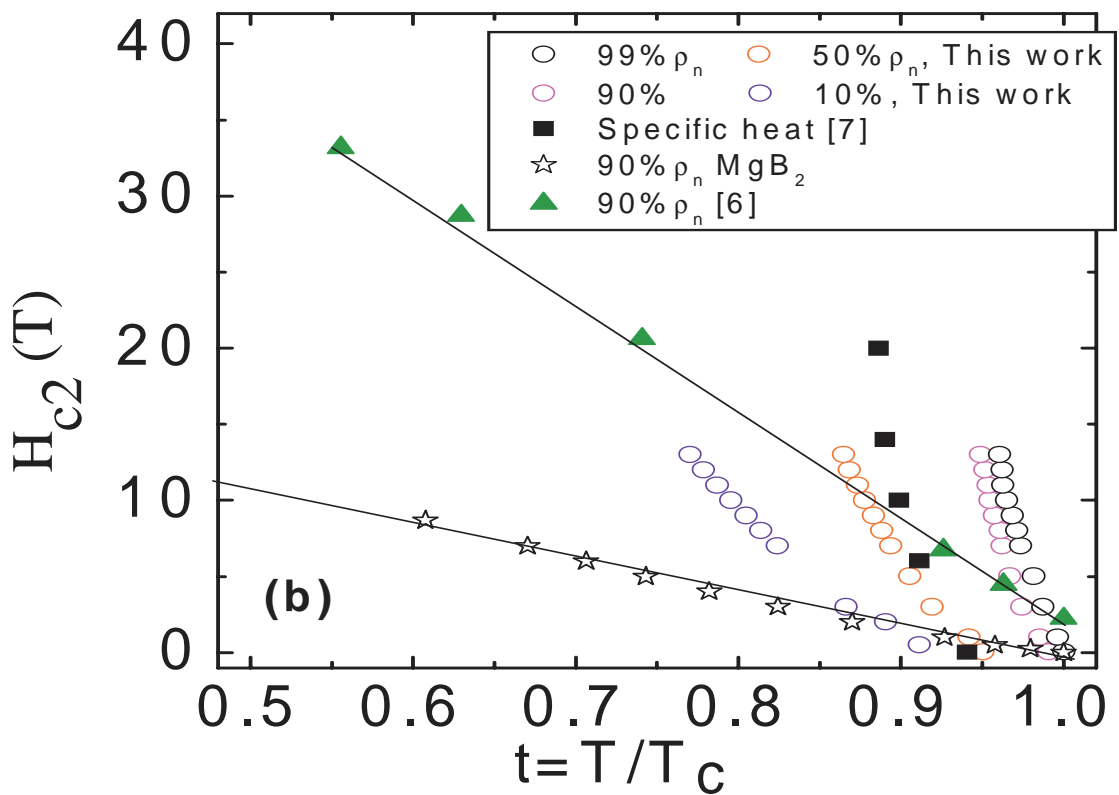

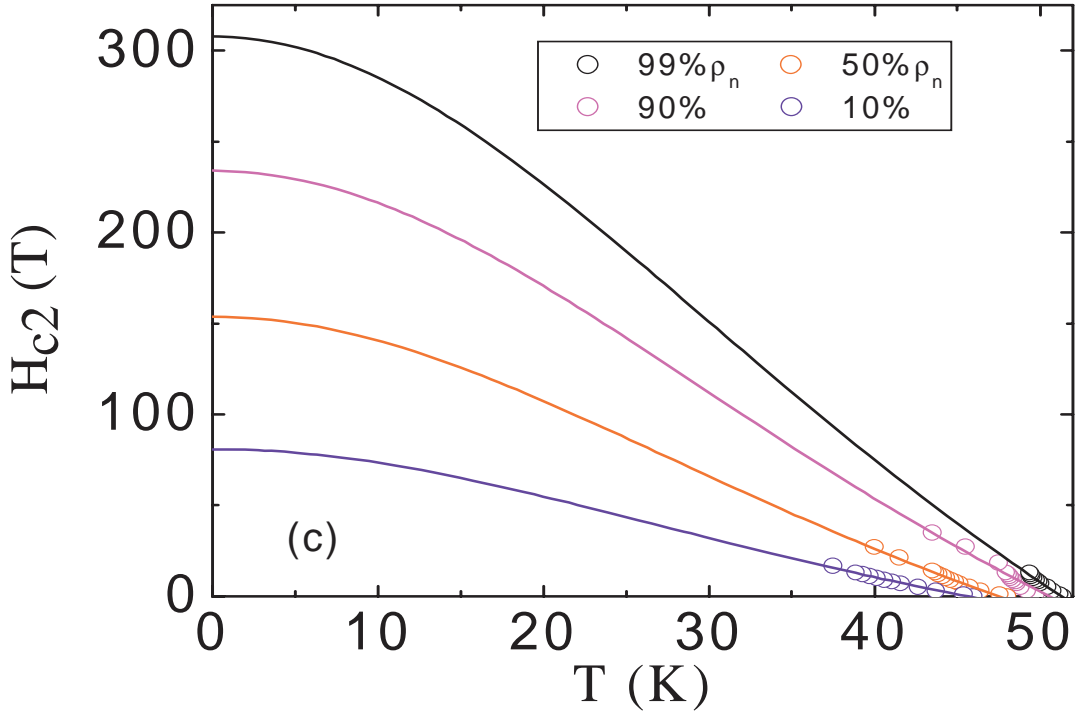

Fig. 4. Field-temperature (a,c) and field-reduced temperature (b) plots derived from measurements of the resistivity against temperature and against magnetic field. $H_{c2}$ is variously defined as the field at which the resistivity drops 99%, 90%, 50%, and 10% from its normal state near $T_c$. The data determined by a 90% drop in resistance for $MgB_2$ and LaFeAsO, and by heat capacity measurements for SmFeAsO are also given. The solid lines in (c) show the theoretical curves based on GL theory (Eq. 1). The solid lines in (b) are only guides to the eyes.

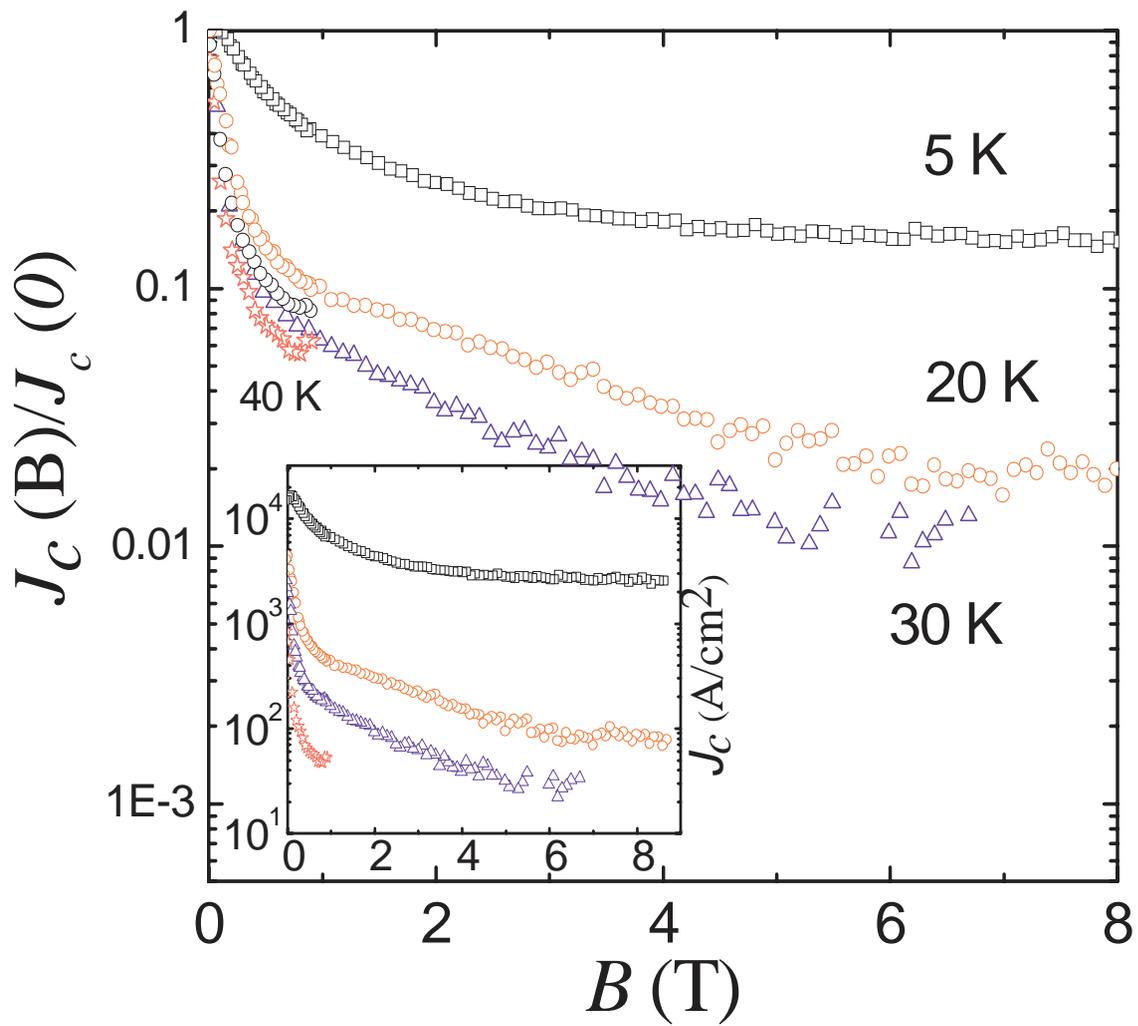

Fig. 5. The normalized $J_c$ vs. field at different temperatures. The inset shows $J_c$ calculated from magnetic measurements based on the real sample size.